\documentclass{article}
\usepackage{spconf,amsmath,graphicx,cite,amsfonts,url,subfig}
\usepackage{algorithm,algorithmic}
\usepackage{bm,multirow}
\title{Time-domain Audio Source Separation based on \\Wave-U-Net Combined with Discrete Wavelet Transform}
\name{Tomohiko Nakamura, \qquad Hiroshi Saruwatari\thanks{This work was supported by JSPS KAKENHI Grant Number JP19H01116.}
}
\address{
    Graduate School of Information Science and Technology, The University of Tokyo\\
    7-3-1 Hongo, Bunkyo-ku, Tokyo 113-8656, Japan
}

\def\R{\mathbb{R}}
\def\R{\mathbb{R}}

\def\even{\mathrm{(even)}}
\def\odd{\mathrm{(odd)}}
\def\fe{f^{\mathrm{(e)}}}
\def\Ce{C^{\mathrm{(e)}}}
\def\fd{f^{\mathrm{(d)}}}
\def\Cd{C^{\mathrm{(d)}}}
\def\Cm{C^{\mathrm{(m)}}}

\def\Cout{C^{\mathrm{(s)}}}

\begin{document}
\ninept

\maketitle
\begin{abstract}
We propose a time-domain audio source separation method using down-sampling (DS) and up-sampling (US) layers based on a discrete wavelet transform (DWT).
The proposed method is based on one of the state-of-the-art deep neural networks, Wave-U-Net, which successively down-samples and up-samples feature maps.
We find that this architecture resembles that of multiresolution analysis, and reveal that the DS layers of Wave-U-Net cause aliasing and may discard information useful for the separation.
Although the effects of these problems may be reduced by training, to achieve a more reliable source separation method, we should design DS layers capable of overcoming the problems.
With this belief, focusing on the fact that the DWT has an anti-aliasing filter and the perfect reconstruction property, we design the proposed layers.
Experiments on music source separation show the efficacy of the proposed method and the importance of simultaneously considering the anti-aliasing filters and the perfect reconstruction property.
\end{abstract}

\section{Introduction}
\label{sec:intro}
Deep neural networks (DNNs) have shown promising results in supervised audio source separation \cite{Sisec2018}.
Most DNN-based methods perform source separation in the magnitude (or power) spectrogram domain \cite{Hershey2016,Jansson2017ISMIR,Takahashi2017ICASSP,Takahashi2018IWAENC}.
However, this approach has several drawbacks.
First, the approach frequently uses the phase of the observed complex spectrogram as that of each separated spectrogram, but the resulting spectrogram may be inconsistent, i.e., no corresponding time-domain signal is guaranteed to exist when we use redundant time-frequency transform \cite{LeRoux2008SAPS}.
Second, although we can use phase reconstruction algorithms such as the Griffin--Lim algorithm \cite{Griffin1984TASSP} to obtain consistent separated spectrograms, these algorithms increase the computation time.
Finally, ignoring the phase in the separation process may lead to suboptimal solutions.
To overcome these drawbacks, an end-to-end approach has been actively explored \cite{Stoller2018ISMIR,Venkataramani2018ACSSC,Wichern2018IWAENC,Slizovskaia2019ICASSP,Luo2019TASLP,Lluis2019Interspeech, Kavalerov2019WASPAA}.

\begin{figure*}[t]
\centering
{
    \hfill
    \subfloat[Multiresolution analysis and synthesis with $L$ levels.]{
        \centering
        \includegraphics[width=0.75\columnwidth]{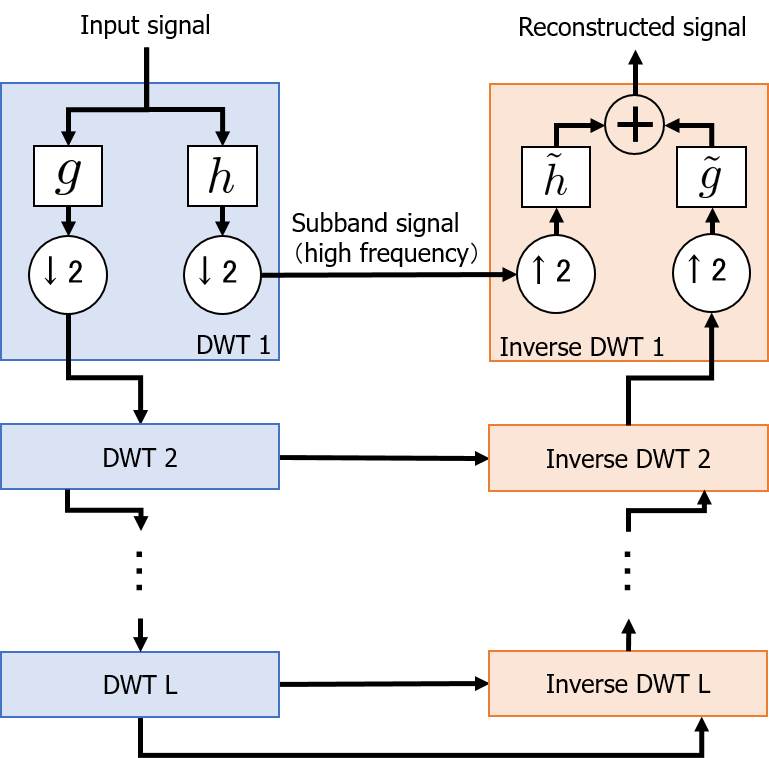}
        \label{fig:MRA}
    }
    \hfill
    \subfloat[Proposed model with $L$ levels and $N$ sources.]{
        \centering
        \includegraphics[width=0.9\columnwidth]{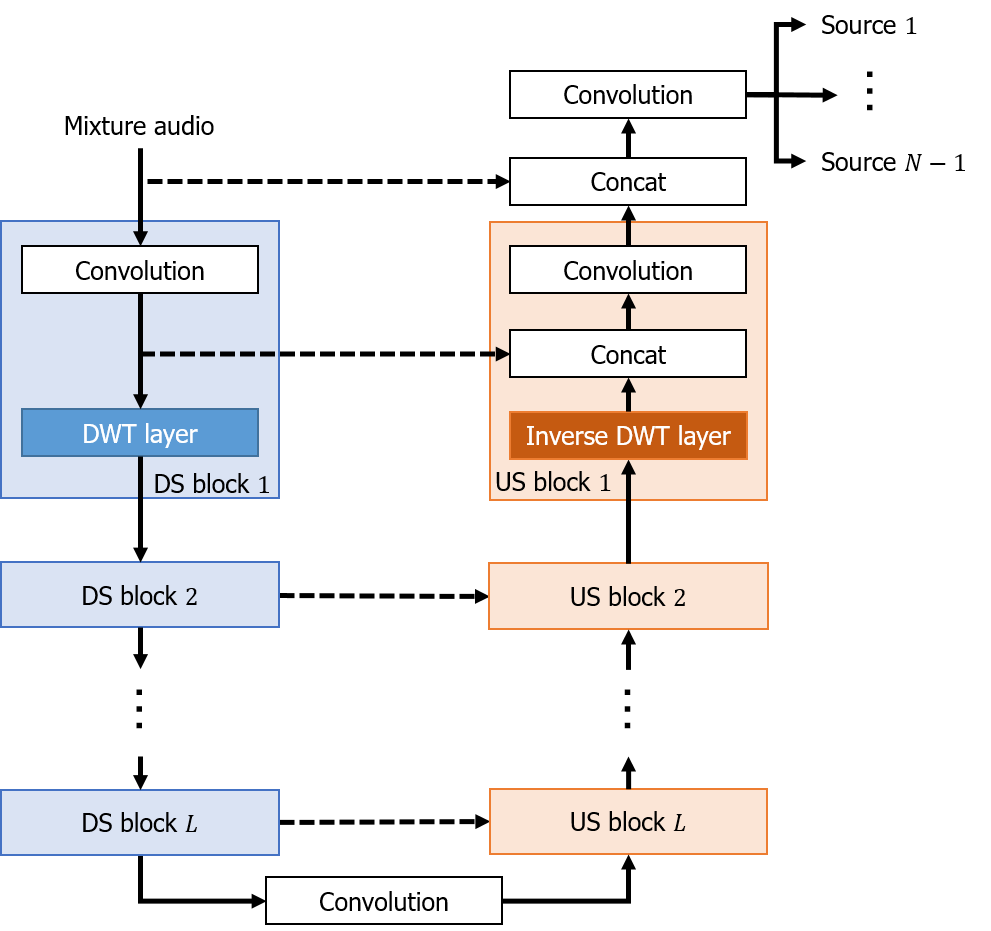}
        \label{fig:Proposed}
    }
    \hfill
}
\caption{Schematic illustration of multiresolution analysis and the proposed model.
The blue and orange rectangles represent down-sampling and up-sampling, respectively.
$h$ and $g$ ($\tilde{h}$ and $\tilde{g}$) denote high-pass and low-pass (reconstruction) filters corresponding to the DWT (inverse DWT).
``Concat" denotes the concatenation of two inputs along the channel axis.
The dashed lines stand for skipped connections.
}
\label{fig:main}
\end{figure*}

One of state-of-the-art end-to-end DNNs is Wave-U-Net \cite{Stoller2018ISMIR}, which directly separates a time-domain signal into source signals.
Wave-U-Net is based on the U-net architecture \cite{Ronneberger2015MICCAI} and has an encoder--decoder architecture.
The encoder successively down-samples feature maps with down-sampling (DS) blocks to halve the time resolution of the feature maps, and the decoder up-samples the feature maps with up-sampling (US) blocks to double the time resolution of the feature maps.
This architecture resembles that of MultiResolution Analysis (MRA) \cite{Mallat1989PAMI} (see Fig.~\ref{fig:main}~\subref{fig:MRA}).
It repeatedly decomposes a signal to subband signals with half the time resolution using a discrete wavelet transform (DWT), and the input signal can be perfectly reconstructed from the subband signals with an inverse DWT.
Roughly speaking, the DS and US blocks correspond to the DWT and inverse DWT, respectively.
This analogy lets us notice the underlying problems of Wave-U-Net.

In the DS blocks, DS is implemented as the decimation layer without low-pass filters, which causes aliasing.
Owing to this phenomenon, for example, shifting an input signal by one sample may change the output of the network markedly and degrade separation performance.
One way to reduce the aliasing artifacts is to insert anti-aliasing filters before DS as in \cite{Mairal2014NIPS,Zhang2019ICML}.
Although this method can be used for our case, another problem remains unsolved.

The decimation layer with or without the anti-aliasing filters is apparently not invertible and discards part of the feature maps even though the discarded components may contain information useful for source separation.
Whether the layer can propagate such information to the following layers strongly depends on training.
To achieve a more reliable source separation method, rather than expecting the model to be trained as we wish, it would be better to design a DS block that can simultaneously overcome the above problems.

With this in mind, we propose novel DS and US layers using the DWT and inverse DWT, and develop an extension of Wave-U-Net combined with the proposed layers (see Fig.~\ref{fig:main}~\subref{fig:Proposed}).
Since a DWT can be seen as a filterbank including a low-pass filter and satisfies the perfect reconstruction property, the proposed layers ensure that the DS blocks have anti-aliasing filters and preserve the entire information of the feature maps.
Note that commonly used DS layers, such as max pooling, average pooling and convolution layers with strides, lacks either anti-aliasing filters or the perfect reconstruction property, and our proposed layers can be useful for many existing DNNs.
The contributions of this paper are summarized as follows:

\noindent \hspace{1mm} 1) We reveal that the conventional DS layers lack either anti-aliasing filters or the perfect reconstruction property through the analogy between the architecture in Wave-U-Net and MRA.

\noindent \hspace{1mm} 2) We propose DS and US layers using the DWT and inverse DWT and develop an extension of Wave-U-Net by incorporating the layers.

\noindent \hspace{1mm} 3) Through experiments on music source separation, we show the efficacy of the proposed model and the importance of simultaneously considering the anti-aliasing filters and the perfect reconstruction property.

\section{Related Works}
\label{sec:related}
The aliasing problem has been investigated in both audio processing \cite{Gong2018Interspeech} and image processing \cite{Zeiler2014ECCV}.
To reduce aliasing artifacts, methods of introducing anti-aliasing filters have been developed for image processing \cite{Mairal2014NIPS,Zhang2019ICML,Williams2018ICLR}.
In \cite{Williams2018ICLR}, a wavelet-based pooling layer has been presented.
The pooling layer outputs only the second-order wavelet subbands of a feature map; therefore, it lacks the perfect reconstruction property.
In contrast, our proposed layers not only include anti-aliasing filters but also satisfy the perfect reconstruction property.
To our knowledge, this study is the first to address this lack of the perfect reconstruction property in DNNs.

As a component of normalizing flow models for image generation, the squeezing operation has been developed to halve the spatial resolution of the feature map without changing the number of elements \cite{Dinh2017ICLR}.
The time-domain adaptation of this operation simply splits the feature map into the odd-sample and even-sample components and concatenates them along the channel axis, which obviously lacks anti-aliasing filters.
Indeed, as shown in section~\ref{sec:proposed_layers}, the operation is part of the process for our DS layer.
This study is the first to reveal the relationship between the operation and the DWT.

\section{Proposed Model}
\label{sec:HWT}
In this section we describe the motivation to derive the proposed layers and develop an extension of Wave-U-Net.

\subsection{Motivation}
\label{sec:motivation}
\begin{figure}[t]
    \centering
    \subfloat[DWT layer.]{
        \includegraphics[width=0.71\columnwidth,clip]{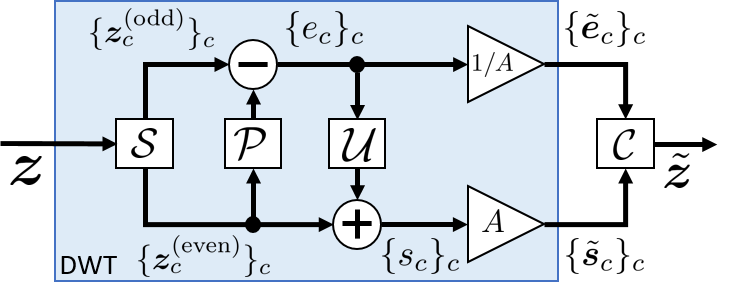}
    }
    \\
    \centering
    \vspace{-2mm}
    \subfloat[Inverse DWT layer.]{
        \includegraphics[width=0.66\columnwidth]{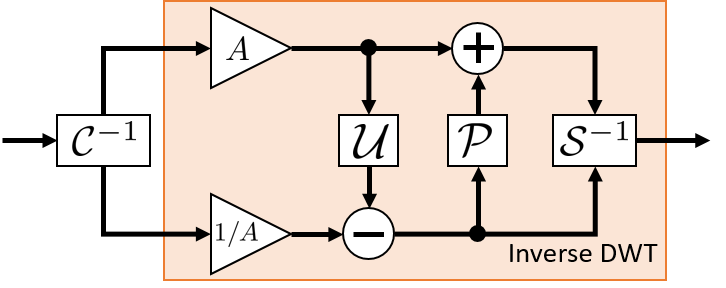}
    }
    \caption{
        Block diagrams of the proposed layers.
        $\mathcal{C}^{-1}$ and $\mathcal{S}^{-1}$ denote the inverse operations of $\mathcal{C}$ and $\mathcal{S}$, respectively.
    }
    \label{fig:proposed_layers}
\end{figure}

As described in section~\ref{sec:intro}, we found an analogy between the architecture in Wave-U-Net and MRA.
This analogy reveals the lack of the anti-aliasing filters and the lack of the perfect reconstruction property.
Although the anti-aliasing problem can be solved by introducing low-pass filters before DS, the other problem remains unsolved.

The decimation layer with and without the anti-aliasing filters discards part of the inputted feature map.
If the layers preceding the decimation layer are trained so that they pack the components useful for the separation into the decimated feature map, the model works well.
If this is not the case, the decoder needs to acquire a way to compensate for the discarded components during the training, which may lead to overfitting.
One may think that, owing to the U-net architecture, each US block is connected to the DS block at the same level of hierarchy and the decoder can access the discarded components.
However, the convolution layer contained in the US block is translation-invariant and it is difficult to identify which elements of the feature map coming through the connections are discarded by the decimation layer.
Thus, whether the decoder compensates for the discarded components strongly depends on training.

In contrast to the decimation layer, the DWT has an anti-aliasing filter and satisfies the perfect reconstruction property, i.e., it can simultaneously overcome the problems.
Focusing on this feature, to achieve a more reliable source separation method, we design novel DS and US layers, called the DWT and inverse DWT layers, respectively, in the following.

\subsection{DWT and Inverse DWT Layers}
\label{sec:proposed_layers}

\begin{table}[t]
\centering
\caption{Detailed architecture of the proposed model}
{\footnotesize
\begin{tabular}{c|c}
    Block & Layer \\ \hline \hline
    & $\mathsf{Input}$ \\ \hline
    \multirow{2}{*}{\begin{tabular}{c}DS block $l$\\ for $l=1,\cdots,L$\end{tabular}}
                & $\mathsf{Conv1D(} \Ce l, \fe \mathsf{)}$\\
                & DWT layer \\ \hline
    Intermediate block & $\mathsf{Conv1D(}\Cm, \fe \mathsf{)}$\\
    \hline
    \multirow{3}{*}{\begin{tabular}{c}US block $l$\\ for $l=L,\cdots,1$\end{tabular}}
                                & Inverse DWT layer \\
                                & $\mathsf{Concat(}\mathsf{DS\ feature}\ l \mathsf{)}$\\
                                & $\mathsf{Conv1D(}\Cd l,\fd \mathsf{)}$ \\
                            \hline
    & $\mathsf{Concat(}\mathsf{Input}\mathsf{)}$ \\
    & $\mathsf{Conv1D(}\Cout(N-1),1 \mathsf{)}$
\end{tabular}
}
\label{tab:arch_details}
\end{table}

Let us consider the feature map $\bm{z}=[\bm{z}_{1}, \cdots, \bm{z}_{C}]\in \R^{T\times C}$, where $T$ and $C$ are the numbers of time points and feature channels, and denote the feature channel index by $c$.
For simplicity, we consider $T$ to be even.
The DWT layer first applies the DWT to $\bm{z}_{c}$ according to the lifting scheme \cite{Sweldens1998SIAM}, which is a computationally efficient technique for the DWT and inverse DWT compared with the Mallat algorithm.
The scheme consists of four steps:
The first is the time-split step in which each feature channel of $\bm{z}_{c}$ is split into the odd-sample and even-sample components, $\bm{z}_{c}^{\odd}\in\R^{T/2}$ and $\bm{z}_{c}^{\even}\in\R^{T/2}$, respectively.
We denote this split operation by $\mathcal{S}$.
The second step is the prediction, in which a prediction of $\bm{z}_{c}^{\odd}$ is computed from $\bm{z}_{c}^{\even}$ using the prediction operator $\mathcal{P}$, and the prediction is subtracted from $\bm{z}_{c}^{\odd}$ to obtain the error component $\bm{e}_{c}\in\R^{T/2}$.
\begin{align}
    \bm{e}_{c} &= \bm{z}_{c}^{\odd} - \mathcal{P}\bm{z}_{c}^{\even}. \label{eq:prediction_step}
\end{align}
This step corresponds to a high-pass filter.
Since $\bm{z}_{c}^{\even}$ contains aliasing artifacts caused by the time-split step, in the third step, called the update step, the smoothed even-sample component $\bm{s}_{c}\in\R^{T/2}$ is computed by applying the update operator $\mathcal{U}$ to $\bm{e}_{c}$ and adding it to $\bm{z}_{c}^{\even}$.
\begin{align}
    \bm{s}_{c} &= \bm{z}_{c}^{\even} + \mathcal{U}\bm{e}_{c}. \label{eq:update_step}
\end{align}
This step corresponds to a low-pass filter.
The final step is the scaling step, in which $\bm{s}_{c}$ and $\bm{e}_{c}$ are scaled by a normalization constant $A$ and its reciprocal, respectively, and we obtain $\tilde{\bm{s}}_{c}=A\bm{s}_{c}$ and $\tilde{\bm{e}}_{c}=\bm{e}_{c}/A$.
After applying the above operations to all channels of $z$, the DWT layer concatenates $\{\tilde{\bm{s}}_{c}\}_{c=1}^{C}$ and $\{\tilde{\bm{e}}_{c}\}_{c=1}^{C}$ along the channel axis to form the down-sampled feature map $\tilde{\bm{z}}\in\R^{T/2 \times 2C}$.
\begin{align}
    \tilde{\bm{z}} = [\tilde{\bm{e}_{1}},\cdots,\tilde{\bm{e}}_{C},\tilde{\bm{s}}_{1},\cdots,\tilde{\bm{s}}_{C}].
\end{align}
We denote the channel concatenation operation by $\mathcal{C}$.
The inverse DWT layer performs the reverse of the above process.
Block diagrams of the two proposed layers are shown in Fig.~\ref{fig:proposed_layers}.

For odd $T$, the time-split step is difficult to directly apply, and inserting a padding layer before the step is required so that the number of time points of the padded feature map equals $T+1$.
We experimentally determined the use of the reflection padding.

The operators $\mathcal{P}$ and $\mathcal{U}$ and the normalization constant $A$ are determined according to the wavelets.
For example, for Haar wavelets, we have
$\mathcal{P}=I_{T}, \mathcal{U}=0.5 \cdot I_{T}$ and $A=\sqrt{2}$ where $I_{T}$ is the $T$ identity matrix.
Note that if we choose the lazy wavelet \cite{Sweldens1996ACHA}, which means that there are no low-pass or high-pass filters, the DWT layer reduces to the time-domain adaptation of the squeezing operation, i.e., $\mathcal{P}=0_{T},\mathcal{U}=0_{T}$ and $A=1$, where $0_{T}$ is the $T$ zero matrix.

Compared with the decimation and linear US layers, the DWT and inverse DWT layers with the Haar wavelets require $TC/2$ subtractions, $TC/2$ additions, and $TC/2$ multiplications for the prediction, update, and scaling steps, respectively.
However, these computations are parallelizable at each step and the processing time does not significantly increase at the DWT and inverse DWT layers.

\subsection{Incorporation of Proposed Layers into Wave-U-Net}
\label{sec:incorporate}
In this section, we incorporate the proposed layers into Wave-U-Net.
A schematic illustration of the proposed model is shown in Fig.~\ref{fig:main}~\subref{fig:Proposed}.
It has an encoder--decoder architecture as in Wave-U-Net\footnote{We adopted the best architecture reported by the authors; M4 \cite{Stoller2018ISMIR}.}.
The encoder (decoder) consists of $L$ DS blocks (US blocks), each of which halves (doubles) the time resolution of the feature maps with the DWT layer (the inverse DWT layer).
Let $l=1,\cdots,L$, $N$ and $\Cout$ be the level index, the number of sources and the number of channels of the input signal, respectively.
The $l$th US block can access the feature map before the DWT layer of the $l$th DS block.
The network outputs $N-1$ source estimates with $\Cout$ channels, and the $N$th source estimate is then obtained by subtracting the sum of the $N-1$ source estimates from the input signal, which ensures that the sum of the source estimates equals the input signal.

The architecture is described in detail in Table~\ref{tab:arch_details}.
$\mathsf{Conv1D(}x,y\mathsf{)}$ denotes a one-dimensional convolution layer with $x$ filters of size $y$.
$\mathsf{Concat(}x\mathsf{)}$ represents the concatenation of the output of the preceding layer and $x$ along the channel axis.
To make the number of time points in the two inputs equal, $x$ is center-cropped before the concatenation.
$\mathsf{Input}$ and $\mathsf{DS\ feature}\ l$ respectively represent the input signal and the feature map before the DWT layer of the $l$th DS block.
All convolution layers are without padding, and the feature maps obtained with the convolution layers can have an odd number of time points.
As described in section~\ref{sec:proposed_layers}, we insert the reflection padding layer before each DWT layer, and discard the last time elements of the feature maps obtained with each inverse DWT layer as in Wave-U-Net.
All convolution layers except for the last are followed by the leaky ReLU, and the last one is followed by the hyperbolic tangent function to normalize the values of the source estimates within $(-1,1)$.
Note that if we replace the DWT and inverse DWT layers with the decimation and linear US layers, respectively, and remove the reflection padding layers, the proposed model is reduced to Wave-U-Net.

\section{Experimental Evaluation}
\label{sec:exp}

\subsection{Experimental Setup}
To evaluate the proposed model, we conducted experiments using the MUSDB18 dataset \cite{musdb18}, which consists of $100$ and $50$ songs for training and test data, respectively.
For all songs, mixture audios and separate recordings of bass, drums, vocals and other instruments are available.
For validation, $25$ songs were randomly selected from the training data, and all recordings were down-sampled to $22.05$ kHz in the stereo format.
We implemented all models by extending the open implementation of Wave-U-Net (\url{https://github.com/f90/Wave-U-Net}) and used the same experimental settings as \cite{Stoller2018ISMIR}.

We compared the proposed model ({\em Proposed}) with the vanilla Wave-U-Net ({\em Wave-U-Net}), which we re-trained, and its two variants ({\em Average Pooling} and {\em Squeezing}).
The variants were used to separately evaluate the effects of the anti-aliasing filters and the perfect reconstruction property.
As DS and US layers, we used the average pooling with a kernel size of $2$ and a stride of $2$ and the linear interpolation layer for {\em Average Pooling} and the squeezing operation and its inverse operation for {\em Squeezing}.

Although the average pooling and decimation layers do not change the channel size of the feature map, the DWT layers and the squeezing operations double that of the feature map.
Owing to these features, it is not easy to exactly match the proposed model and Wave-U-Net in terms of the number of parameters.
For this reason, we also used a variant of {\em Proposed} with a reduced number of parameters ({\em Proposed (small)}) and another variant of Wave-U-Net ({\em Wave-U-Net+}) with an increased number of parameters to allow a fair comparison.
We set $\Ce=12$ for {\em Proposed (small)}, $\Ce=24$ for {\em Proposed}, {\em Squeezing} and {\em Wave-U-Net} and $\Ce=48$ for {\em Wave-U-Net+} and {\em Average Pooling}.
The characteristics of all models are summarized in Table~\ref{tab:compare_methods}.
Following the best architecture of Wave-U-Net reported by the authors \cite{Stoller2018ISMIR}, we used stereo inputs, i.e., $\Cout=2$, and set $L=12$, $\Cm=312$, $\Cd=24$, $\fe=15$ and $\fd=5$ for all models.
For the DWT and inverse DWT layers, we used Haar wavelets.

During training, we randomly cropped audio segments of $147443$ samples from the full audio signals and multiplied random gains within $[0.7,1.0]$ as a data augmentation to form a batch with size $16$.
The loss function was the mean squared error function over all elements in the batch, and the Adam optimizer with a learning rate of $0.0001$ and decay rates of $\beta_1=0.9$ and $\beta_2=0.999$ was employed.
We continued to train each model until no improvements of the validation loss were observed during $20$ successive epochs.
With the same stopping criterion, we then fine-tuned the model with the learning rate of $0.00001$ and the batch size of $32$, and selected the model at the epoch with the lowest validation loss.

\begin{table}[t]
\centering
\caption{Characteristics of all models. ``perf. rec. prop." means the perfect reconstruction property}
{
\footnotesize
    \begin{tabular}{c||c|ccc}
    \multirow{2}{*}{Model} & \multirow{2}{*}{\#params} & Has & Satisfies \\
    & & anti-aliasing filters & perf. rec. prop.
    \\ \hline \hline
        {Wave-U-Net+} & $28.31$M & No&No \\
        {Wave-U-Net} & $10.26$M & No&No \\ \hline
        {Proposed} & $15.15$M & Yes&Yes \\
        {Proposed (small)} & $5.81$M & Yes&Yes \\
        {Average Pooling} & $28.31$M & Yes&No \\
        {Squeezing} & $15.51$M & No&Yes
    \end{tabular}
}
\label{tab:compare_methods}
\end{table}

\subsection{Results}
\begin{figure}[t]
    \centering
    \includegraphics[width=0.95\columnwidth,clip]{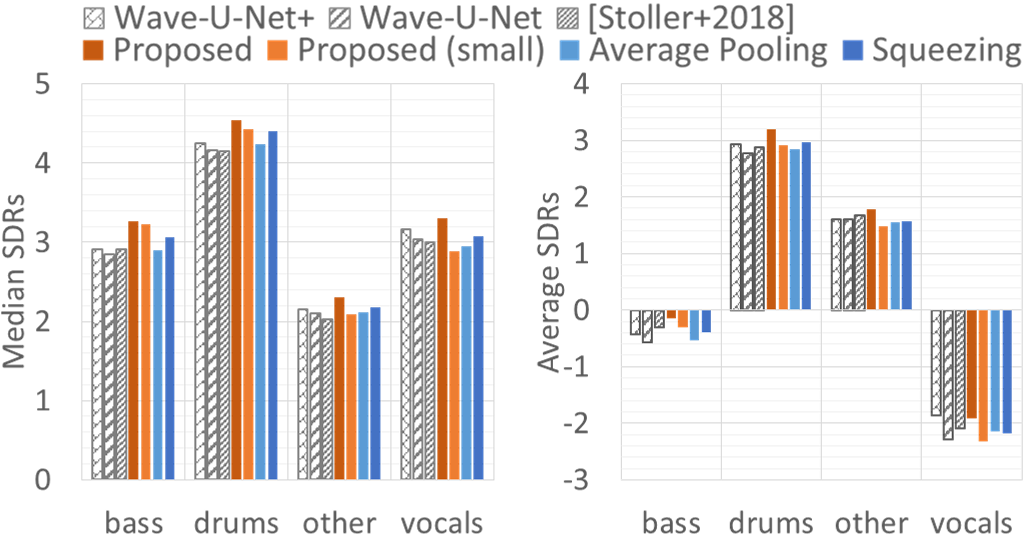}
    \caption{
        Median and average SDRs of all models. The values are averaged over five trials.
    }
    \label{fig:results}
\end{figure}

Fig.~\ref{fig:results} shows the median and average source-to-distortion ratios (SDRs) for all models, which were calculated using the SiSEC2018 evaluation procedure \cite{Sisec2018} and averaged over five trials.
Note that [Stoller+2018] denotes the results reported in \cite{Stoller2018ISMIR}.
{\em Proposed} outperformed {\em Wave-U-Net} and [Stoller+2018] for all musical instruments and achieved a (slightly) higher performance for bass and drums (vocals and other) compared with {\em Wave-U-Net+} despite the fact that {\em Proposed} has only half the number of parameters as {\em Wave-U-Net+}.
This clearly shows the efficacy of the proposed model.
Furthermore, even with half the number of parameters as {\em Wave-U-Net}, {\em Proposed (small)} provided higher and competitive median and average SDRs compared with the conventional models for all the instruments except for vocals.
We can confirm that the proposed layers can better capture the important features and are more suitable for audio source separation than the conventional DS and US layers used in Wave-U-Net.
While we used the Haar wavelet in the experiments, the use of other wavelets may further improve separation performance, and we will examine, as a future work, which wavelets are suitable.

{\em Average Pooling} and {\em Squeezing} did not reach the performance of {\em Proposed}.
We observed that {\em Squeezing} provided lower median signal-to-interference ratios than {\em Proposed} and higher median signal-to-artifact ratios than {\em Average Pooling} relatively.
This implies that reducing the aliasing artifacts makes it easier to distinguish the target source from other sources and introducing the perfect reconstruction property makes it easier to propagate the entire information of the signal through the network.
These results show that both the anti-aliasing filters and the perfect reconstruction property should be simultaneously considered when designing the neural network.

\begin{figure}[t]
    \centering
    \includegraphics[width=0.95\columnwidth,clip]{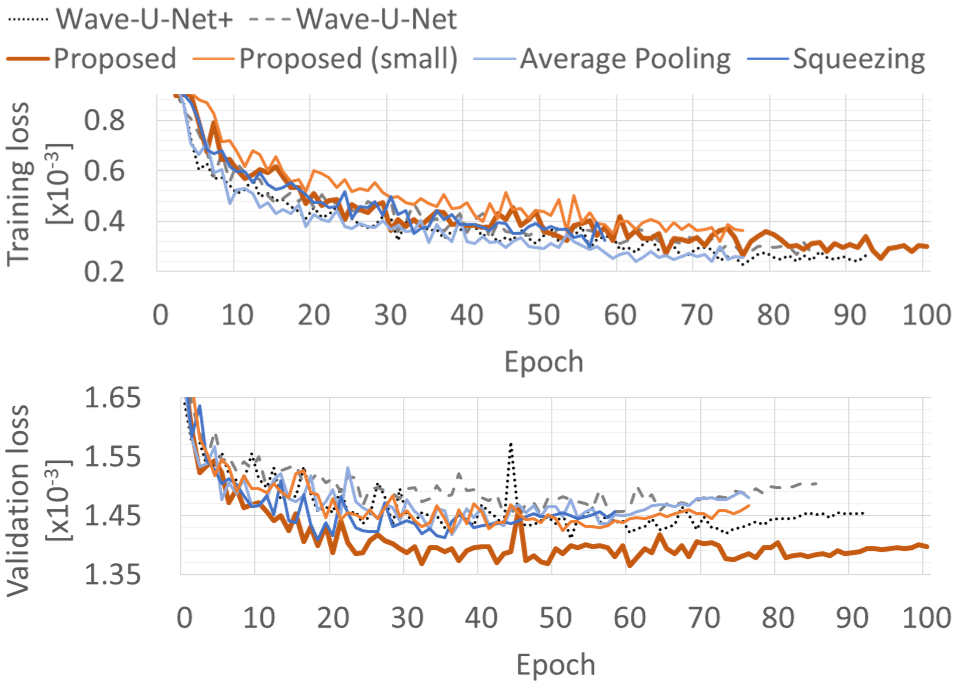}
    \caption{Training and validation losses averaged per epoch.}
    \label{fig:curves}
\end{figure}

As shown in Fig.~\ref{fig:curves}, we found that {\em Proposed} showed greater training losses but smaller validation losses than the other models at most of the training epochs.
This suggests that the proposed layers can reduce overfitting, but its further investigation is required.

\section{Conclusion}
We presented an extension of Wave-U-Net combined with the DWT and inverse DWT layers.
The motivation to develop the proposed layers came from our observation that the architecture of Wave-U-Net resembles that of MRA.
Through this analogy, we found the two problems of the DS layers in Wave-U-Net: the lack of the anti-aliasing filters and the lack of the perfect reconstruction property.
To simultaneously overcome these problems, we designed the DWT layer.
The experiments on music source separation show the efficacy of the proposed model and the importance of simultaneously considering the aliasing filters and the perfect reconstruction property.

\vfill\pagebreak

\bibliographystyle{IEEEbib}
\bibliography{refs}

\end{document}